\documentclass[proof]{pasj01}
\draft

\begin{document} 
\Received{}%{yyyy/mm/dd}
\Accepted{}%{yyyy/mm/dd}
%\Published{yyyy/mm/dd}

\title{Orbital solution leading to an acceptable interpretation for the enigmatic gamma-ray binary $\mathbf{HESS~J0632+057}$}

%%% begin:list of authors
% Do NOT capitalize all letters in "textsc".
\author{Yuki \textsc{Moritani}\altaffilmark{1}%
\thanks{E-mail: yuki.moritani@ipmu.jp}}
\altaffiltext{1}{Kavli Institute for the Physics and Mathematics of the Universe (WPI), \\
   The University of Tokyo Institutes for Advanced Study, The University of Tokyo, \\
 5-1-5, Kashiwa, Chiba 277-8583, Japan}
%\email{yuki.moritani@ipmu.jp}

\author{Takafumi \textsc{Kawano}\altaffilmark{2}
\thanks{alumnus}}
\altaffiltext{2}{Department of Physical Science, Hiroshima University, \\
1-3-1 Kagamiyama, Higashi-Hiroshima 739-8526, Japan}
%\email{ccccc@xxx.xxx.xx.xx}

\author{Sho \textsc{Chimasu}\altaffilmark{3}
\thanks{alumnus}}

\author{Akiko \textsc{Kawachi}\altaffilmark{3}}
\altaffiltext{3}{Department of Physics, School of Science, Tokai University, \\
Hiratsuka, Kanagawa 259-1292, Japan}
%\email{bbbbb@xxx.xxx.xx.xx}

\author{Hiromitsu \textsc{Takahashi}\altaffilmark{2}}

\author{Jumpei \textsc{Takata}\altaffilmark{4}}
\altaffiltext{4}{School of Physics, Huazhong University of Science and Technology, \\
Wuhan, 430074, China}

\author{Alex C. \textsc{Carciofi}\altaffilmark{5}}
\altaffiltext{5}{Instituto de Astronomia, Geof\'{i}sica e Ci\^{e}ncias Atmosf\'{e}ricas, Universidade de S\~{a}o Paulo, \\
Rua do Mat\~{a}o 1226, Cidade Universit\'{a}ria, S\~{a}o Paulo, SP 05508-090, Brazil}

%%% end:list of authors

%% `\KeyWords{}' always has to be placed before `\maketitle'.
\KeyWords{binaries: general --- stars: emission-line, Be --- stars: individual: ($\mathrm{HESS~J0632+057}$)} %Do NOT move this preamble from here!

\maketitle

\begin{abstract}
% 300 words (298) 
High-dispersion spectroscopic monitoring of $\mathrm{HESS~J0632+057}$ has been carried out over four orbital cycles in order to search for orbital modulation, covering the entire orbital phase.
We have measured radial velocity of H$\alpha$ emission line with the method introduced by \citet{Shafter1986}, which has been successfully applied to some Be stars.
The velocity is seen to increase much earlier than expected for the orbital period of 315 days, and much more steeply than expected at around ``apastron".
The period of the H$\alpha$ modulation is found to be as $308^{+26}_{-23}$ days.
We have also analyzed Swift/XRT data from 2009 to 2015 to study the orbital modulation, selecting the data with good statistics ($\geq$ 30 counts).
With additional two-year data to the previous works, the orbital period has been updated to $313^{+11}_{-8}$ days, which is consistent with the previous X-ray periods and the spectroscopic one.
The past XMM-Newton and Chandra observations prefer the period of 313 days.
With the new period, assuming that H$\alpha$ velocities accurately trace the motion of the Be star, we have derived a new set of the orbital parameters.
In the new orbit, which is less eccentric ($e \sim 0.6$), two outbursts occur after apastron, and just after periastron.
Besides, the column density in bright phase ($4.7^{+0.9}_{-08}\times10^{21}\;\mathrm{cm^{-2}}$) is higher than in faint phase ($2.2\pm0.5\times10^{21}\;\mathrm{cm^{-2}}$).
These facts suggest that outbursts occur when the compact object passes nearby/through the Be disk.
The mass function implies that mass of the compact object is less than 2.5 $\mathrm{M_{sun}}$ assuming that the mass of the Be star is 13.2--18.2 $\mathrm{M_{sun}}$ \citep{Aragona2010} unless the inclination is extremely small.
The photon index indicates that the spectra becomes softer when the system is bright.
These suggest that the compact object is a pulsar.
\end{abstract}

%------------------------
\section{Introduction}\label{sec:intro}
$\mathrm{HESS~J0632+057}$, an X-ray binary hosting a Be star, is identified as a gamma-ray binary \citep{Aharonian2007}.
Gamma-ray binaries, literally, have an enhanced emission in gamma-rays ($>$ 1 MeV), and show multi-wavelength variabilities up to TeV energies with orbital modulation [see \citet{Dubus2013} for a recent review].
At present, 6 systems are established as gamma-ray binaries including one in the LMC \citep{Corbet2016}.
In most systems the nature of the compact object is under discussion, which makes it difficult to understand the mechanism of particle acceleration and hence such high-energy emissions.
Besides, combined with the fact that only a few such systems among more than 100 High-Mass X-ray Binaries show enhanced gamma-ray emissions in the Milky Way, the veiled nature of the compact objects makes it difficult to understand what kind of evolution track is related to the gamma-ray binaries.

Among the gamma-ray binaries, $\mathrm{HESS~J0632+057}$ shows puzzling variabilities for its orbit of $P_\mathrm{{orb}}=315-321$ days \citep{Bongiorno2011,Aliu2014} and $e=0.837$ \citep{Casares2012}.
In X-rays, the system is very faint around apastron.
More interestingly, outburst occurs both before and after the very faint phase.
The former outburst, called primary outburst, has sharp rise and decay, while the latter, called secondary outburst, has a long duration (several tens of days) and flat-shaped peak.
The orbit indicates that the compact object is far from the Be star around apastron ($\sim 100 \; \mathrm{R_{Be}}$, where $\mathrm{R_{Be}}$ is the equatorial radius of the Be star).
With such a distance, it is difficult to explain the sudden and periodic change in brightness with a simple scenario, in which the X-ray emission around the compact object is associated with the density of the surrounding region, because the Be disk should have low density and not be bended.
Very high-energy emissions ($>$TeV) is rather in phase with X-ray emissions \citep[for example]{Aliu2014}.
It was very recent that GeV emission was detected in $\mathrm{HESS~J0632+057}$ by deeply searching the Fermi nine-year data \citep{Li2017}.

In the optical, \citet{Aragona2010,Moritani2015} reported ``S-shaped" variability in H$\alpha$ line profile.
Both variations were detected after apastron, and \citet{Moritani2015} found that the variability had terminated around periastron.
Because ``S-shaped" variability is thought to originate from perturbation in the structure of the Be disk, it is difficult to explain what excites such a variation at apastron.

In addition to these puzzling activities, the lack of established geometry also makes $\mathrm{HESS~J0632+057}$ an enigmatic source.
Because the proposed orbital periods, which are derived from X-ray light curve, are different from each other by $\sim$1 week [321 days by \citet{Bongiorno2011}, 315 days by \citet{Aliu2014}], the predicted time of periastron deviates by $\sim$ 1 month now, several years after the discovery.
The orbital period has not been estimated using optical data, because there has not been enough observation to cover more than one orbital cycle.

Recently, \citet{Moritani2015} reported that short-term variability occurred $\sim$ 1 month after apastron, but no significant variation at periastron.
These features imply small tidal interaction between the disk and the compact object, and the X-ray light curve would be explained in the framework of ``flip-flop" pulsar scenario, which was originally proposed for another gamma-ray binary LS~I$+$60$^{\circ}$~303 \citep{Torres2012}.
In this scenario, the pulsar wind is quenched by gas pressure of the dense Be disk around periastron.
However, for confirmation of this scenario, further observations are required around the primary outburst, when the pulsar wind is predicted to recover and start pushing the Be disk again.

We have continued to monitor $\mathrm{HESS~J0632+057}$ to obtain a complete phase coverage, partly motivated to detect the line profile variability expected at the primary outburst.
Four-year monitoring has almost completed the phase coverage and the data are not biased at given phases.

In this paper, we discuss the variability in the radial velocity of H$\alpha$ and obtain an improved orbit.
\citet{Casares2012} reported that the centroid velocity of H$\alpha$ varied more rapidly at apastron than at periastron.
Similar trend at apastron was shown by \citet{Moritani2015}.
However, the emission lines originate from the Be disk, so that these variations can be associated with disk activity.
To mitigate the effect of Be-disk activity, the bisector velocity of H$\alpha$ wing has been measured following the method by \citet{Shafter1986}.

Besides, we have analyzed Swift/XRT archived data to study orbital modulation in X-rays and compare it with the optical data.
Because there have been two-year more data available since \citet{Aliu2014}, we searched the periodicity again.
The periods from the optical and X-ray data are compared.

%------------------------
\section{Observations} \label{sec:obs}
\subsection{Optical}
We have monitored $\mathrm{HESS~J0632+057}$ using high dispersion spectrographs for four years to cover the entire orbital phase --- season 1: 14 nights from 2013 October to 2014 April, season 2: 17 nights from 2014 November to 2015 April, season 3: 6 nights from 2016 January to March, and season 4: 9 nights from 2016 December to 2017 April.
We have already published profiles from season 1 \citep{Moritani2015}, when the profile variability occurred around apastron.

The observations were executed using mainly the 188 cm telescope and HIDES (High Dispersion Echelle Spectrograph) with fiber-fed system \citep{Kambe2013} at Okayama Astrophysical Observatory (OAO).
HIDES covers 4200--7400 \AA \ wavelength range, with typical wavelength resolution $R$ of $\sim 50000$.
A few spectra were obtained by ESPaDOnS (Echelle SpectroPolarimetric Device for the Observation of Stars) at the Canada France Hawaii Telescope \citep{Manset(2003)}.
The wavelength coverage  is 3700--10500 \AA \ and the resolving power is  $\sim68000$.

The OAO/HIDES data were reduced in the standard way, using IRAF\footnote{http://iraf.noao.edu/} {\tt echelle} package --- subtraction of bias, flat fielding, extracting spectra, calibration of the wavelength using Th-Ar lines, normalization of the continuum, and helio-centric correction of the radial velocity.
For CFHT/ESPaDOnS data, on the other hand, we have obtained reduced data using Libre-Esprit/Upena\footnote{http://www.cfht.hawaii.edu/Instruments/Upena/} pipeline, provided by the instrument team, and have rectified the normalized intensity, in order to compare with OAO/HIDES data. 

\begin{figure}
 \begin{center}
  \includegraphics[width=8cm]{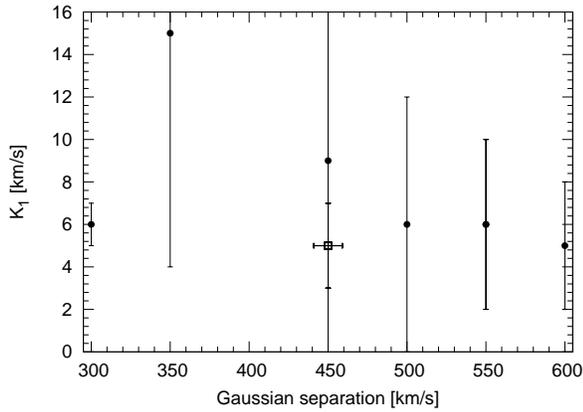} 
 \end{center}
\caption{
The projected amplitude of the radial velocity $K_1$ as a function of Gaussian separation.
Orbital parameters are estimated by fixing the orbital period at 313 days (see section \ref{sec:res_x}).
Square indicates the average Gaussian separation ($450\pm9 \;\mathrm{km\;s^{-1}}$) used for this work, and $K_1$ ($5\pm2 \;\mathrm{km\;s^{-1}}$, see Table \ref{tab:orbit}) calculated using the resultant radial velocity.
Circles indicate $K_1$ using the radial velocity measured by fixing the Gaussian separation among the profiles.
}\label{fig:disg}
\end{figure}

The radial velocity has been measured for the H$\alpha$ emission line.
Here we have adopted the method introduced by \citet{Shafter1986}.
This method is suitable to measure radial velocity as bisector velocity of the emission profile wing, and has been successfully applied to some Be stars \citep[and reference therein]{Dulaney2017}.
The typical error of the measurement is 0.5 $\mathrm{km\;s^{-1}}$.
The measurement may depend on the selection of Gaussian separation, which is the distance between the peaks of the Gaussian kernels.
We have arbitrarily set the separation for each individual profile as the width of 25\% of the peak intensity, resulting in 450$\pm$9 $\mathrm{km\;s^{-1}}$ on average among the profiles.
To check the feasibility of this prescription, we have compared the projected amplitude of the radial velocity ($K_1$, see below) with that derived by fixing the separation from 300 $\mathrm{km\;s^{-1}}$ to 600 $\mathrm{km\;s^{-1}}$ in 50 $\mathrm{km\;s^{-1}}$ step.
Figure \ref{fig:disg} shows the evolution of $K_1$ as a function of the Gaussian separation.
$K_1$ seems to be constant when the separation is fixed between 500 $\mathrm{km\;s^{-1}}$ and 600 $\mathrm{km\;s^{-1}}$.
Judging from the  fact that constant value (6  $\mathrm{km\;s^{-1}}$) is consistent with the case where the separation is changed per profile (square in Figure \ref{fig:disg}), the measurement should reasonably eliminate the effect of the Be disk activity.

%% --- Figure : Diagnostic diagram --- %%
%% --- Figure : Diagnostic diagram (end) --- %%

\subsection{X-ray}
We have analyzed public data of the Swift X-ray Telescope (XRT) obtained from 2009 January 26th to 2015 January 30th.
XRT sensitivity ranges from 0.3--10 keV bands \citep{Burrows2005}.
Since the previous work \citep{Aliu2014}, $\mathrm{HESS~J0632+057}$ has been observed by Swift/XRT for about two years.
Among the 7-year data, we have selected bright 191 data points which have good statistics ($\geq$ 30 counts).
The typical duration of these data is 4 -- 5 ks.
The rejected data show low quality either because exposure was too short, or because $\mathrm{HESS~J0632+057}$ was faint at other phases than the two outbursts.

For data reduction, we have utilized the {\tt HEAsoft}\footnote{https://heasarc.nasa.gov/lheasoft/} 6.13 package which includes {\tt XSPEC} version 12.8.0, and the calibration files version 2015/07/21.

In order to investigate spectral variation associated with brightness, we set a threshold of 0.04~counts~s$^{-1}$ for the entire energy band (horizontal lines in Figure \ref{fig:x-lc}). 
Hereafter we call events which have the count rate above (below) the threshold as bright (faint) events.
There are 21 bright events, which occur around peak of two outbursts.
Note that we also have set the thresholds to 0.03 and 0.05 counts s$^{-1}$, and obtained consistent results.

%------------------------
\section{Results} \label{sec:result}
\subsection{Optical}\label{sec:result-opt}

%% --- Figure : radial velocity --- %%
\begin{figure*}
 \begin{center}
  \includegraphics[width=5.5cm]{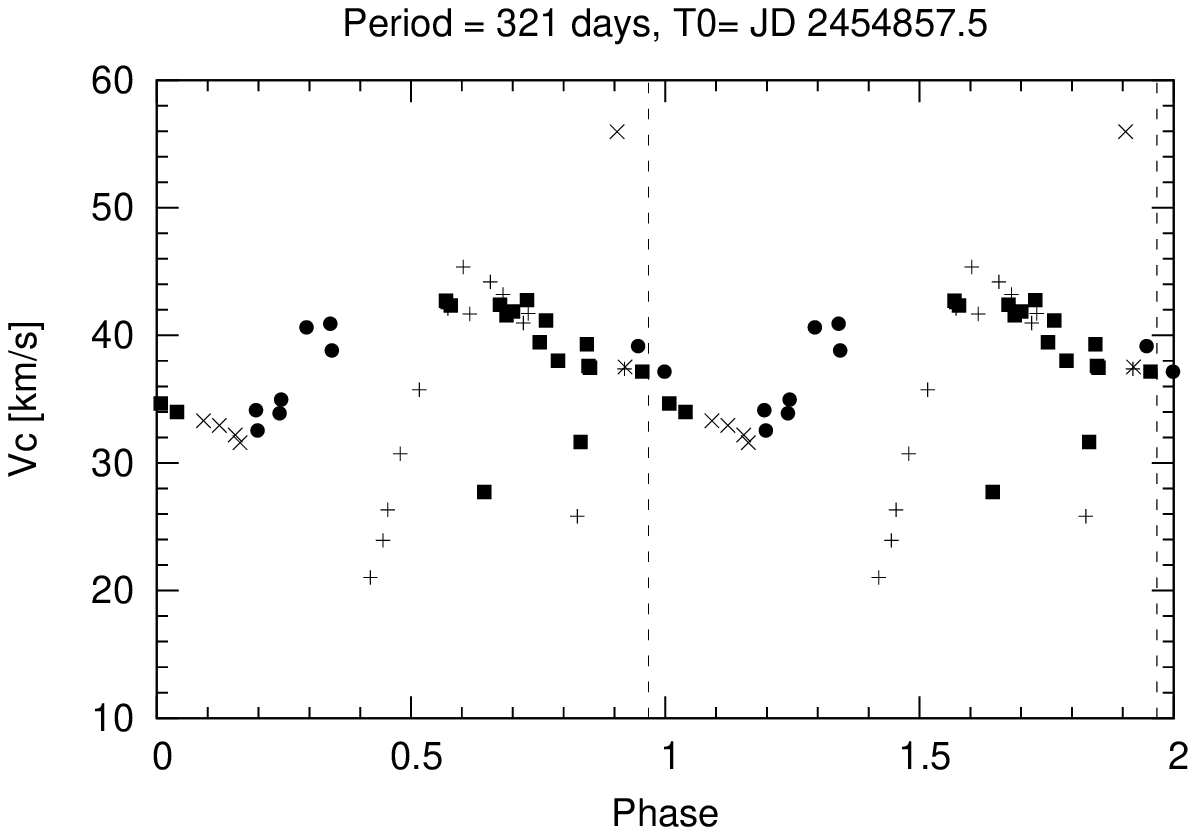} 
  \includegraphics[width=5.5cm]{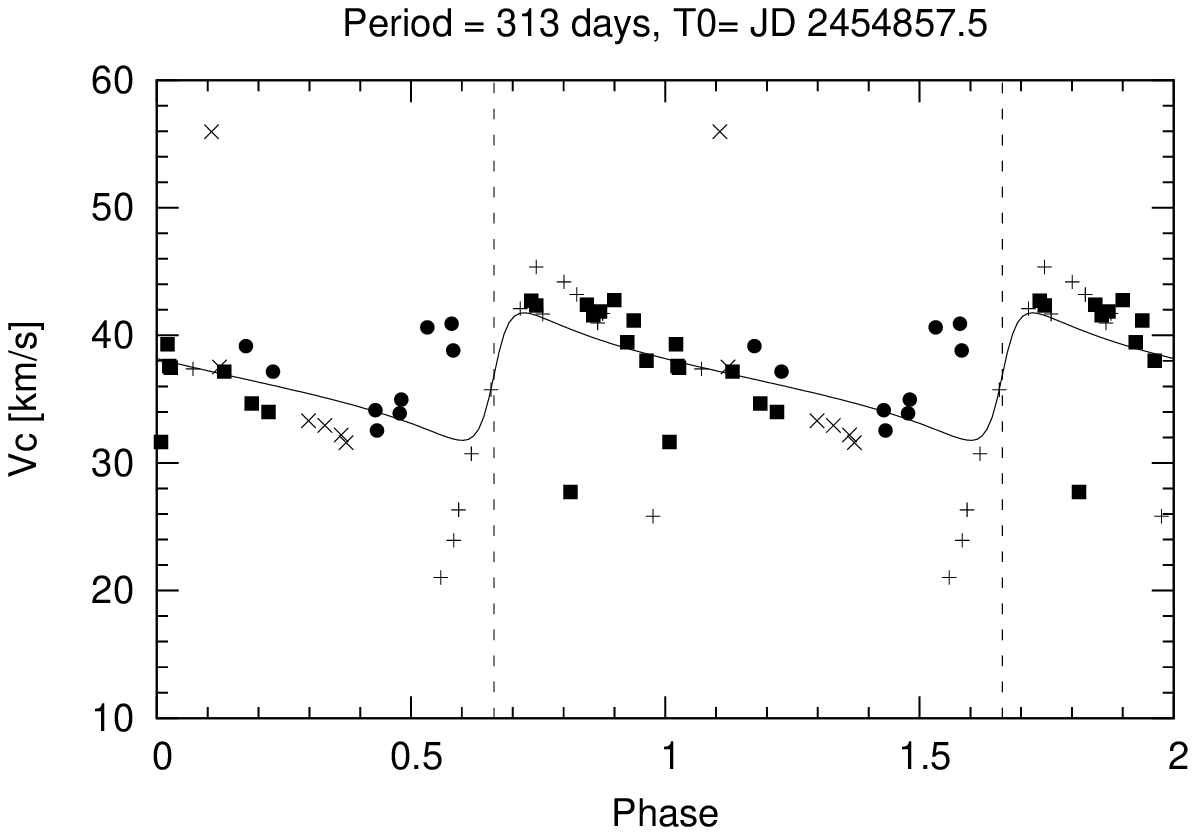} 
  \includegraphics[width=5.5cm]{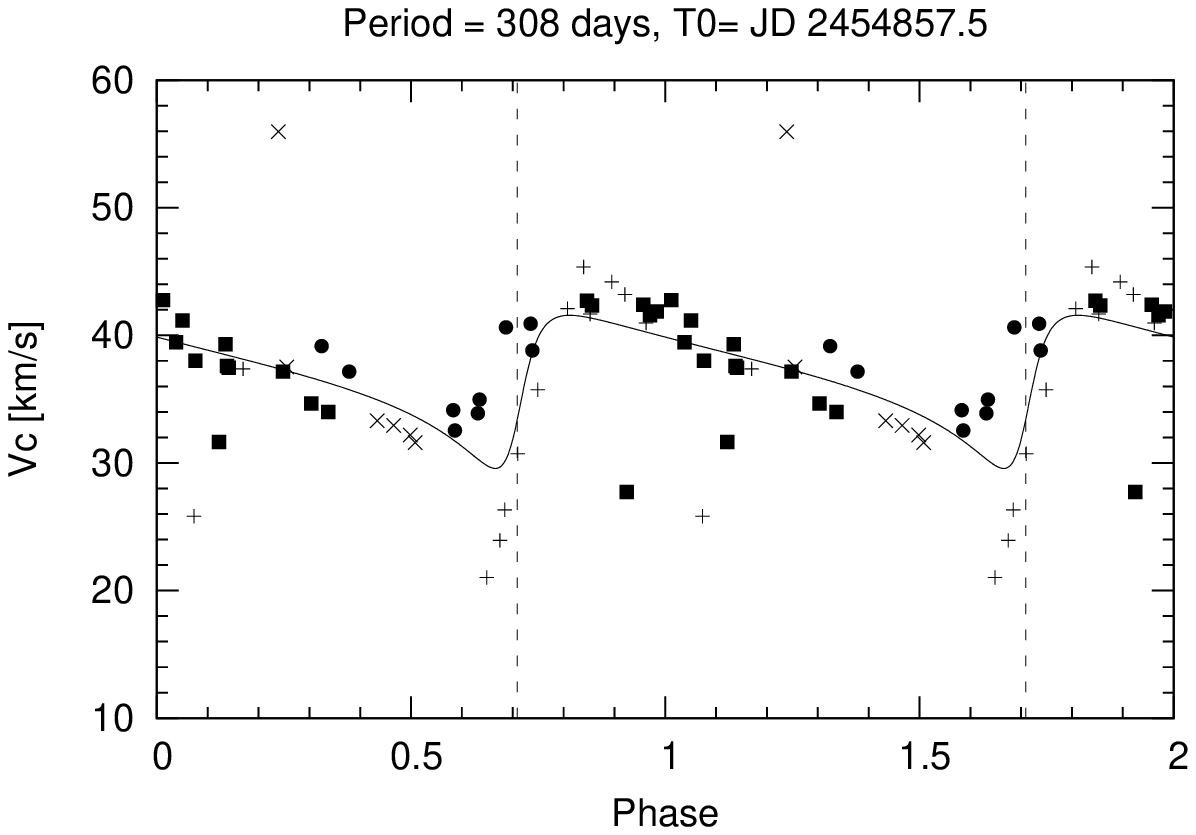} 
 \end{center}
\caption{The radial velocity folded with period of 321 days (left), 313 days (center) and 308 days (right).
Different marks indicate individual observational seasons --- pluses; from 2013 October to 2014 April, squares; from 2014 November to 2015 April, crosses; from 2016 January to March, and circles; from 2016 December to 2017 April.
The phase 0 is cited from \citet{Bongiorno2011} (JD 2454857.5).
Vertical dashed lines indicate estimated periastron and solid line in the center and right panels is modeled curve using the new orbital parameters (Table \ref{tab:orbit}).}\label{fig:radvel}
\end{figure*}
%% --- Figure : radial velocity (End) --- %%

In the left panel of Fig. \ref{fig:radvel}, radial velocity of the H$\alpha$ profile is plotted, folded with the period of 321 days \citep{Bongiorno2011}.
Here the zero phase is arbitrary, cited from \citet{Bongiorno2011}, following the previous works.
Different marks indicate individual observational seasons.
The radial velocity has increased much earlier than expected for a 321-day cycle -- it started to increase around phase ($\phi_{321}$) of 0.2.
Also the radial velocity increases from around $\phi_{321} \sim$ 0.2--0.6, more steeply than it decreases around $\phi_{321} \sim$ 0.6--0.2.
These features are also seen if the velocity is folded with 315 days \citep{Aliu2014}: the velocity experienced more rapid increase at $\phi_{315} \sim$ 0.3--0.6, and slower decrease at $\phi_{315} \sim$ 0.6--0.3.
[See the center panel of Fig. \ref{fig:radvel}, for the period of 313 days (see below) as a reference.
Although the period is different, the features are quite similar to the case of 315 days.]
Indeed, a similar trend was reported by \citet{Casares2012}.

These facts suggest that the orbital period is shorter than predicted by X-ray activity \citep{Aliu2014,Bongiorno2011}, and that the periastron is on the opposite side of the orbit proposed by \citet{Casares2012}.

\begin{figure}
 \begin{center}
  \includegraphics[width=8cm]{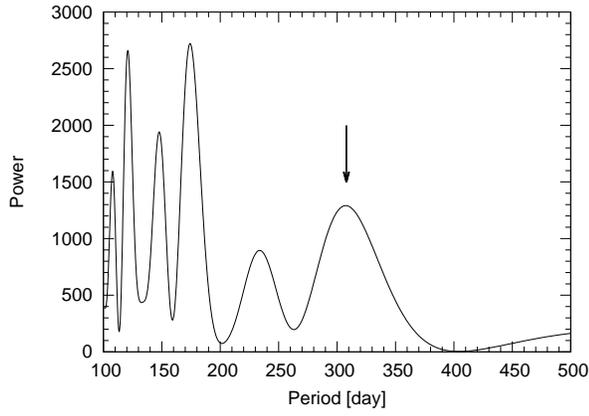} 
 \end{center}
\caption{Power spectrum of Fourier analysis applied to the H$\alpha$ radial velocities.
The unit of the abscissa is converted to days to see the orbital period easily.
The arrow corresponds to 308 days.
}\label{fig:pow_emi}
\end{figure}

Regarding the former possibility, we have applied Fourier analysis to the H$\alpha$ radial velocities.
The resultant period is $308^{+26}_{-23}$ days (Figure \ref{fig:pow_emi}).
Here, the errors have been calculated as the width between the frequency for the peak and 2/$\pi$ of the peak.
The peak is significant in 95 \% confidence, according to the $\chi ^2$ test.
The right panel in Fig. \ref{fig:radvel} shows the same radial velocity, but folded with the period of 308 days.
Slightly but clearly, the turning point from decrease to increase looks smoother in the case of $P=308$ days.

%% --- Figure : Fourier periodgram --- %%
%% --- Figure : Fourier periodgram (end) --- %%

In addition, we have estimated the radial velocity of absorption lines (HeI$\lambda 4471, 5040, 6678$) using the cross-correlation method \citep{Simkin1974}.
Here the spectra were binned to increase the S/N, and masked at the center of HeI$\lambda6678$ lines, where the emission component is visible.
Considering the good phase coverage, the reference spectrum for the correlation has been taken as the average of the all spectra.
Because the resultant velocities have large measurement errors ($\gtrsim 10 \mathrm{km\;s^{-1}}$) and large scattering (standard deviation of $30 \mathrm{km\;s^{-1}}$), we have not been able to obtain significant peaks by Fourier analysis.

To search for the orbital parameters, we have applied the {\tt Binary Star Combined Solution Package} provided by VLT/CHARA team\footnote{http://www.astro.gsu.edu/\~{ }gudehus/binary.html} to the radial velocity, fixing the orbital period at 308 days.
Table \ref{tab:orbit} shows the results, comparing with \citet{Casares2012}.
The orbit predicted in this work has slightly lower eccentricity ($e \sim 0.6$), and smaller semi-major axis ($a_1\sin i \sim 0.14$ AU).
The argument of periastron is also quite different ($\omega=249$ degree).
Note that if the orbital period is set as free parameter, the resultant period was derived to be 304$\pm$3 days, and other parameters for this period are the same within the error as those for 308 days.
Note also that $e$ is the same within the error as \citet{Casares2012}, if we fix the orbital period to 313 days (see the next section).

%% --- Table : Orbital parameters --- %%
\begin{table}
  \tbl{Orbital Parameters of $\mathrm{HESS~J0632+057}$.}{%
  \begin{tabular}{cccc}
      \hline
	Parameter & \citep{Casares2012} &  \multicolumn{2}{c}{This Work} \\
      \hline
	$P_{\mathrm{orb}} \mathrm{[day]}$ & 321$^\mathrm{a}$ & 308$^\mathrm{b}$ & 313$^\mathrm{c}$  \\
	$T_{\mathrm{peri}} \mathrm{[day]}$	& 2 455 167.907	& 2 455 076$\pm$10	& 2 455 065$\pm$11 \\
	$\phi_{\mathrm{peri}}$$^\mathrm{d}$	& 0.967	& 0.709	& 0.663 \\
	$e$  & 0.83$\pm$0.08 & 0.62$\pm$0.16 & 0.64$\pm$0.29  \\
	$\omega \mathrm{[deg]}$ & 129$\pm$17 & 249$\pm$26 & 271$\pm$29  \\
	$K_1 \mathrm{[km \; s^{-1}]}$ & 22.0$\pm$ 5.7 & 6$\pm$1 & 5$\pm$2  \\
	$\gamma \mathrm{[km\;s^{-1}]}$ & 48.3$\pm$8.9	& 36.9$\pm$0.8	& 36.7$\pm$0.9  \\
	$a_1\sin i \mathrm{[AU]}$ & 0.362$\pm$0.261	& 0.136$\pm$0.029	& 0.120$\pm$0.029  \\
	$f \mathrm{[M_{sun}]}$ & $0.06^{+0.15}_{-0.05}$	& 0.0035$\pm$0.0022	& 0.0024$\pm$0.0017 \\
       \hline
    \end{tabular}}\label{tab:orbit}
\begin{tabnote}
a: Cited from \citet{Bongiorno2011}. \\
b: Fixed to the result of the radial velocity. \\
c: Fixed to the result of X-ray light curve. \\
d: The phase at periastron if we follow \citet{Bongiorno2011} for $\phi=0$.
\end{tabnote}
\end{table}
%% --- Table : Orbital parameters (end) --- %%

%% --- Figure : Mass function --- %%
\begin{figure}
 \begin{center}
  \includegraphics[width=8cm]{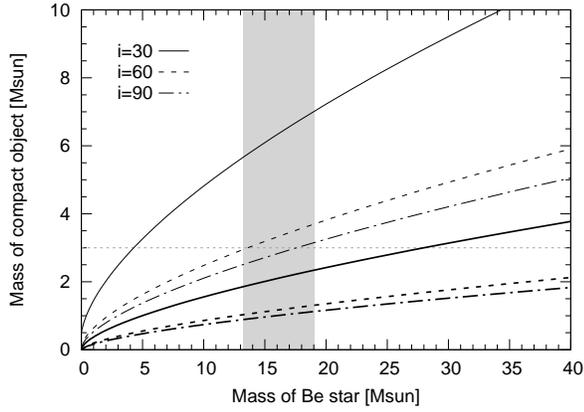} 
 \end{center}
\caption{Mass function of $\mathrm{HESS~J0632+057}$.
Thick and thin lines are results of this work and \citet{Casares2012}, respectively.
Solid, dotted and dashed-dotted lines are the case of inclination angle of 30$^\circ$, 60$^\circ$, and 90$^\circ$, respectively.
Gray rectangle marks proposed mass range of the Be star \citep{Aragona2010}.
Horizontal dotted line denotes Tolman-Oppenheimer-Volkoff limit.}\label{fig:massf}
\end{figure}
%% --- Figure : Mass function (end) --- %%

Figure \ref{fig:massf} displays mass functions of $\mathrm{HESS~J0632+057}$, comparing the result by \citet{Casares2012}.
If the H$\alpha$ velocities trace the motion of the Be star, smaller amplitude of the radial velocity ($K_1=6\;\mathrm{km\;s^{-1}}$) suggests much smaller mass function ( $f=0.0035 \pm 0.0022 \; \mathrm{M_{sun}}$), as shown in Table \ref{tab:orbit}.
\citet{Aragona2010} derived the mass of Be star as 13.2--18.2 $\mathrm{M_{sun}}$.
As a companion of this Be star, the mass of the compact object should be less than 2.5 $\mathrm{M_{sun}}$, unless the inclination angle is very small.

\subsection{X-ray}\label{sec:res_x}

%% --- Figure : X-ray light curve --- %%
\begin{figure*}
 \begin{center}
  \includegraphics[width=8cm]{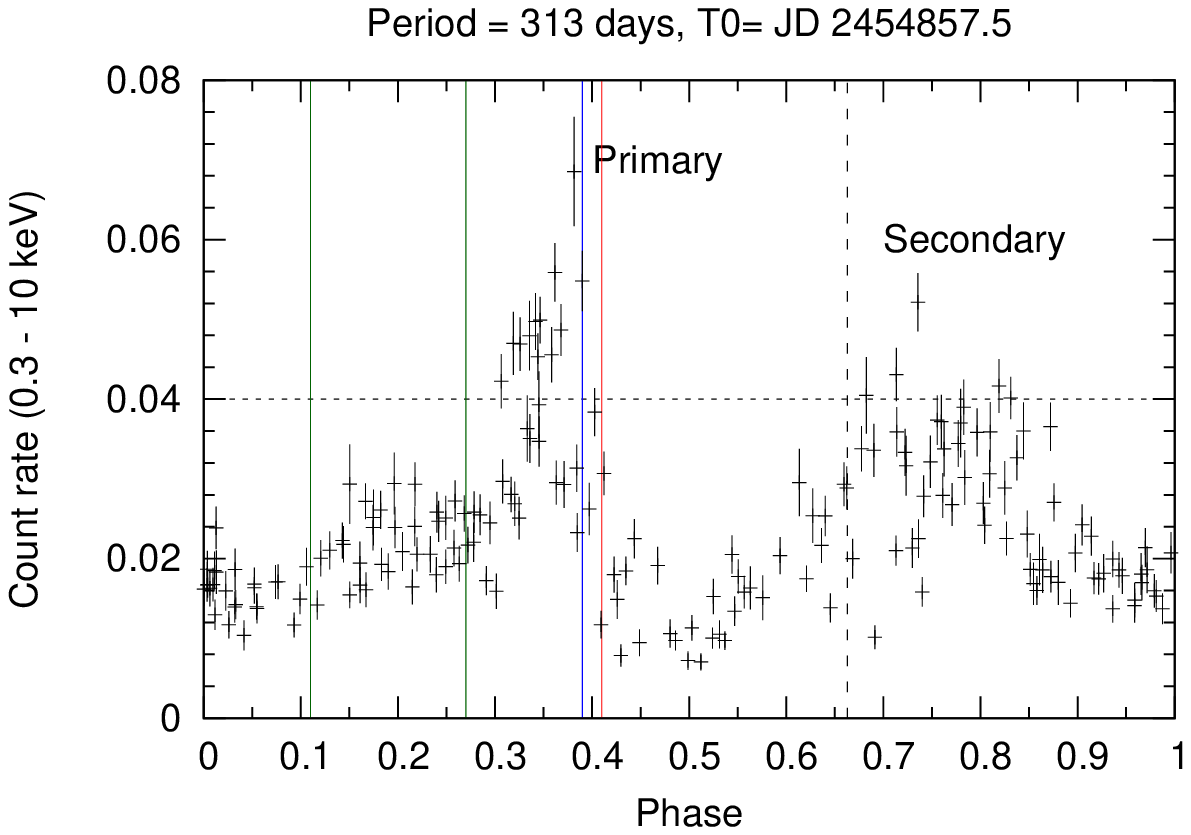} 
  \includegraphics[width=8cm]{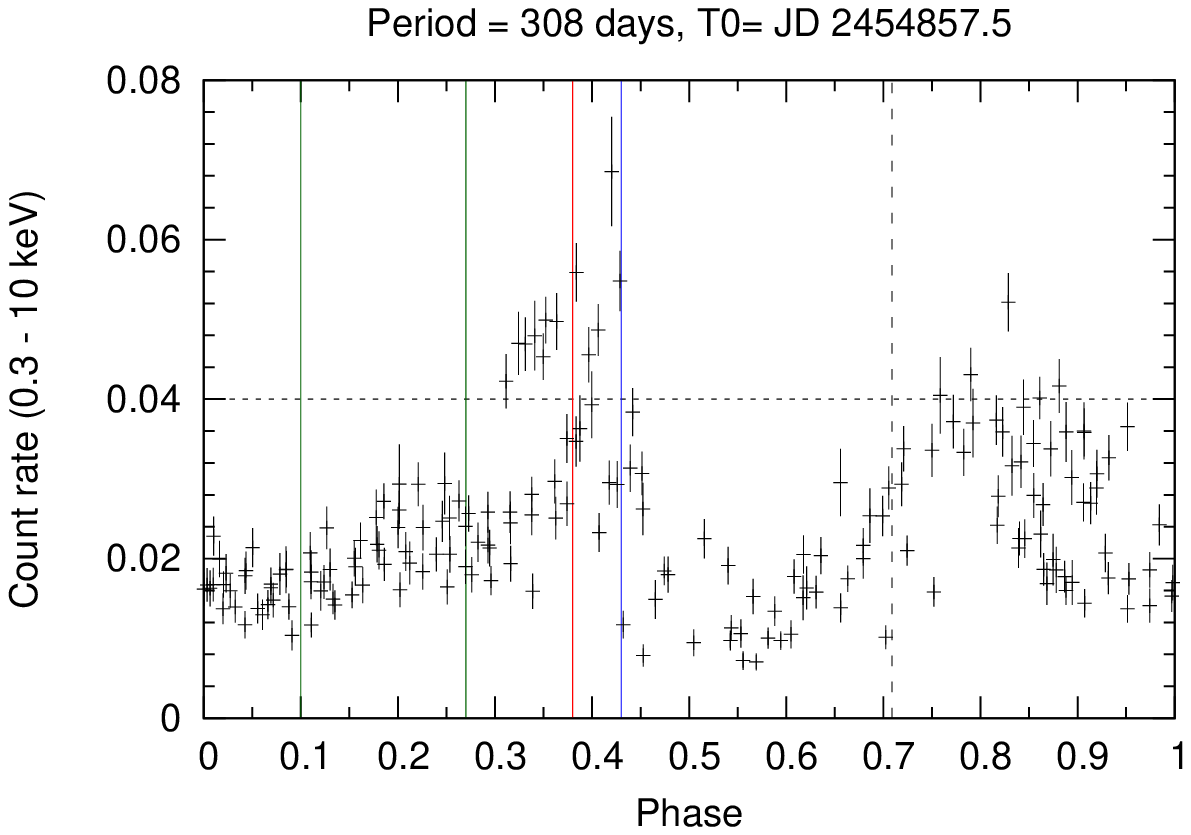} 
 \end{center}
\caption{XRT light curves of $\mathrm{HESS~J0632+057}$, folded with the period of 313 days (left) and 308 days (right).
The count rate is used because of low statistics.
Error bars correspond to 1~$\sigma$ error.
Horizontal dotted line, corresponding to 0.04 ~counts~s$^{-1}$, is the threshold for distinguishing bright and faint events.
Vertical black dashed lines are estimated periastrons, as in Figure \ref{fig:radvel}.
The color vertical lines indicate the phase at the previous X-ray observations (Table \ref{tab:x-spec})--- XMM-Newton (red), Chandra (blue), and Suzaku (green).}\label{fig:x-lc}
\end{figure*}
%% --- Figure : X-ray light curve (end) --- %%

Independently of the optical data, we have searched for periodicity using the XRT light curve.
We have utilized the Z-DCF method \citep{Edelson1988,Alexander1997}, following \citet{ Bongiorno2011} and \citet{Aliu2014}.
The resultant period is $313^{+11}_{-8}$ days, and consistent with \citet{Bongiorno2011} ($321 \pm 5$ days) and \citet{Aliu2014} ($315^{+6}_{-4}$ days).
Note that the error bar of this work is larger than the previous study, in spite of the increased number of data.
This is likely caused by the low cadence of the new data, coupled with possible cycle-to-cycle variability.
The orbital parameters for this period are almost the same as those for 308 days (Table \ref{tab:orbit}).

Figure \ref{fig:x-lc} shows the folded light curve with the orbital period of 313 days (left) and 308 days (right).
Because of poor statistics of Swift/XRT data, we have used the count rates to make the light curves. 

%% --- Figure : X-ray photon index and column density --- %%
\begin{figure}
 \begin{center}
  \includegraphics[width=8cm]{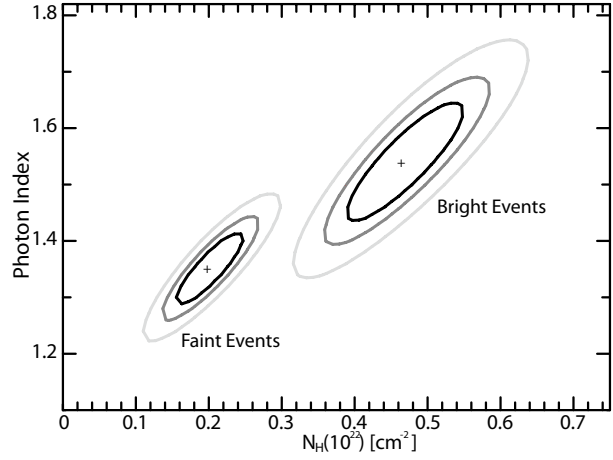} 
 \end{center}
\caption{
Contour plot of the resultant column density $N_\mathrm{H}$ and photon index $\Gamma$.
The confidence levels are 68\% (black), 90\% (dark-gray), and 99\% (light gray), respectively.
Cross marks are the center of the contours: ($N_\mathrm{H}$, $\Gamma$)$=$ (4.7, 1.54) for the bright events and (2.2,1.37) for the faint events.}\label{fig:x-contour}
\end{figure}
%% --- Figure : X-ray photon index and column density (end) --- %%

To analyze the Swift/XRT spectrum, we have re-binned w.r.t. the energy band so that each spectral bin has at least 10 counts.
We have fitted the spectra with an absorbed power-law model ({\tt wabs} and {\tt powerlaw} models in the {\tt XPSEC} v12.8.0).
The column density $N_\mathrm{H}$, the photon index $\Gamma$ and the normalization have been taken as free parameters.
Table \ref{tab:x-spec} lists the resultant parameters. 

Figure~\ref{fig:x-contour} shows contour plots of $N_\mathrm{H}$ and $\Gamma$ for the result.
The contours of the bright and the faint events are centered at ($N_\mathrm{H}$, $\Gamma$)$=$ (4.7, 1.54) and (2.2, 1.37), respectively.
In the bright events, $\mathrm{HESS~J0632+057}$ has higher column density than in the faint events with $>$99\% confidence level.
$N_\mathrm{H}$ in the XRT bright and faint events is consistent with XMM-Newton and Chandra observations, respectively (see Table \ref{tab:x-spec}).
The photon index in the bright events is softer than the faint events, although the difference is less significant.

%% --- Table : Orbital parameters --- %%
\begin{table*}
  \tbl{Summary of column density $N_\mathrm{H}$ and photon index $\Gamma$ dependency of $\mathrm{HESS~J0632+057}$.
See text for details about Swift/XRT bright and faint events. }{%
  \begin{tabular}{cccccccc}
      \hline
	Observatory & Time	& \multicolumn{3}{c}{Phase $\phi ^{a}$} &  $N_\mathrm{H}$  	& $\Gamma$	& flux \\
	&	[YYYY/MM/DD]	& $315$ days	& $313$ days	& $308$ days	& [$10^{21} \; \mathrm{cm^{-2}}$]	& 	& [$10^{-12} \; \mathrm{erg\;s^{-1}\;cm^{-2}}$]\\
      \hline
	XMM-Newton$^{b}$	& 2007/09/01	& 0.42	& 0.41	& 0.38	& $2.1\pm0.4$	& $1.18\pm0.08$	& 0.51$\pm$0.09 (0.3--10 keV)  \\
	Suzaku$^{c}$	& 2008/04/23	& 0.12	& 0.11	& 0.10	& $3.1\pm0.4$	& $1.55\pm0.05$	& 0.94$\pm$0.03 (1--5 keV) \\
	Suzaku$^{c}$	& 2009/04/20	& 0.27	& 0.27	& 0.27	& $2.6\pm0.3$	& $1.38\pm0.03$	& 0.88$\pm$0.02 (1--5 keV) \\
	Chandra$^{d}$	& 2011/02/13	&  0.37	&  0.39	& 0.43	& 4.3$\pm$0.2	& $1.61\pm0.03$	& 3.2$\pm$0.2 (0.3--10 keV)  \\
	Swift/XRT bright	& 2009/01--2015/01	& ---	& ---	& ---	& $4.7_{-0.8}^{+0.9}$	& $1.54_{-0.11}^{+0.12}$	& $>$0.04 [cnt/s]  \\
	Swift/XRT faint		& 2009/01--2015/01	& ---	& ---	&---	& $2.2\pm0.5$	& $1.37\pm0.07$	& $\leq$0.04 [cnt/s] \\
       \hline
    \end{tabular}}\label{tab:x-spec}
\begin{tabnote}
a: $\phi = 0$ is set at JD 2454857.5 to compare with X-ray light curves (Figure \ref{fig:x-lc}). \\
References: (b) \citet{Hinton2009, Rea2011} ; 
(c) \citet{Skilton2011} ;
(d) \citet{Rea2011} 
\end{tabnote}
\end{table*}
%% --- Table : Orbital parameters (end) --- %%

\section{Discussions} \label{sec:discuss}
\subsection{The Updated Orbital Period}
As described above, the radial velocity of H$\alpha$ line increased rapidly around $\phi_{315}$=0.5, and decreased slowly afterwards (Figure \ref{fig:radvel}).
This variation suggests that the periastron and apastron phases are inverted to those derived using absorption lines \citep{Casares2012}.
In fact, \citet{Casares2012} reported the variation in H$\alpha$ which is similar to this work (see their Fig. 4).
It might be possible that the orbital parameters in \citet{Casares2012} were affected by ``cadence" of observation, considering the incomplete coverage of orbital cycle, that is, the slightly biased data around the periastron, and rather large scattering of the data at other phases than the periastron.

The seven-year X-ray light curve has revised  the orbital period as $313^{+11}_{-8}$ days.
Furthermore, the Fourier analysis of the H$\alpha$ radial velocities yields a comparable period ($308^{+26}_{-23}$ days).
These periods are slightly smaller than reported in the previous works ($\geq$315 days) \citep{Bongiorno2011,Aliu2014} although fully consistent within errors.

Table \ref{tab:x-spec} compares the phases at the X-ray observations based on the proposed orbital periods.
\citet{Rea2011} compared $N_\mathrm{H}$ and $\Gamma$ between bright (Chandra) and faint (XMM-Newton) state and found that these parameters are significantly higher when $\mathrm{HESS~J0632+057}$ is bright.
XRT spectra (Figure \ref{fig:x-contour}) show a consistent behavior.
Although XRT observations started two years after the system was observed by XMM-Newton, repeatability of the primary outburst suggests that the light curve before 2009 should be similar to those in Figure  \ref{fig:x-lc}.
Assuming this, if the shortest period (308 days) is adopted, the light curve (right panel of Figure  \ref{fig:x-lc}) indicates that $\mathrm{HESS~J0632+057}$ was brighter during the XMM-Newton observation than during the Chandra observation.
This result is inconsistent with the observed flux \citep{Rea2011} and the relation between the flux and the spectral parameters ($N_\mathrm{H}$ and $\Gamma$).
If 313 days is adopted, on the other hand, the revised phases during the XMM-Newton and Chandra observations are reasonable in terms of the X-ray phenomena.
Therefore, 308 days is likely too short for the updated orbital period of $\mathrm{HESS~J0632+057}$.
On the other hand, the folded radial velocity curve (Figure \ref{fig:radvel}) suggests that 321 days is too long as the orbital period.

\subsection{The New Orbit of $\mathrm{HESS~J0632+057}$ and Interpretations of Phenomena}
\begin{figure}
 \begin{center}
  \includegraphics[width=8cm]{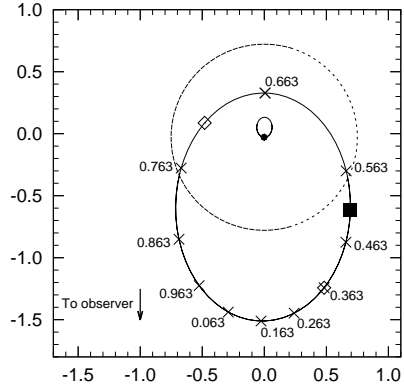} 
 \end{center}
\caption{New orbit of $\mathrm{HESS~J0632+057}$. 
Large and small ellipses in solid line are the orbit of the compact object and the Be star, respectively. 
Here, the mass and radius of the Be star are assumed as  16 $\mathrm{M_{sun}}$ and 8 $\mathrm{R_{sun}}$, the average values in \citet{Casares2012}.
The mass of the compact object, on the other hand, is assumed as 1.4  $\mathrm{M_{sun}}$.
The scale is unit of semi-major axis with assumption of these masses.
For the orbit of the compact object, the phase is marked with crosses with a step of 0.1, following the zero point by the previous works.
Filled circle around the center indicates the Be star at periastron, and dotted circle is Be disk with the radius of 30 $\mathrm{R_{Be}}$ \citep{Moritani2015} in the coplanar case.
Open diamonds, and filled square mark the peak of the X-ray outbursts and X-ray dip.
}\label{fig:orbit}
\end{figure}

The X-ray data indicate that the orbital period is close to 313 days.
The orbital parameters, obtained by fixing the period at 313 days are almost the same as for the 308-day period, except for the semi major axis and the mass function.
Assuming that the mass of the Be star and the compact object is 16 $\mathrm{M_{sun}}$ and 1.4  $\mathrm{M_{sun}}$, respectively, we have drawn the suggested orbit in Figure \ref{fig:orbit} with marks for the X-ray activities.

In the new orbit, the sharp, primary X-ray outburst occurs after apastron (phase of $\sim 0.35$, anomaly of 0.56), while the broad, secondary X-ray outburst just after periastron (phase of $\sim 0.7$, anomaly of 0.22).
These phases (diamonds in Figure \ref{fig:orbit}), combined with the fact that the column density is higher in the bright event, in particular around the peak of the primary outburst (see below), bring a possibility that the outbursts occur when the compact object approaches to the misaligned Be disk, as in another gamma-ray binary ($\mathrm{PSR~B1259-63}$; \cite{Negueruela2011}).
The orbital phase of the primary outburst suggests the Be disk is azimuthally tilted by $\sim 20-30 ^{\circ}$. 
The broadness of the secondary outburst is possibly because of the long time for the compact object to pass through the Be disk around periastron.
The X-ray flux of the secondary outburst, however, is still difficult to understand, because the compact object passes denser part than the primary outburst.

In fact, if we analyze the bright events during the primary and the secondary outbursts separately, the former shows higher $N_\mathrm{H}$ than the faint event, whereas the latter shows a similar $N_\mathrm{H}$ to the faint event.
Note that if we separate the data with respect to the phase, there is no significant feature in the spectra partly due to low statistics.
Further studies including deep X-ray observations around the secondary outburst are required to confirm the new geometry.

The system, on the other hand, is very faint in X-rays at the phase of $\sim 0.5$, where the compact object locates close to the Be disk in the coplanar case.
However, assuming the XMM-Newton spectra is fit with single power-law model, smaller $N_\mathrm{H}$ around this phase suggests that the compact object is not surrounded by the dense region.
This implies that the Be disk tilts so that its upper-right side in Figure \ref{fig:orbit} goes behind the paper, and hence the compact object locates between the observer and the Be disk.
Note that, it cannot be ruled out that the spectra can be described with a broken power-law model, where additional absorption component is assumed.
In such a case, $N_\mathrm{H}$ is expected to increase at very faint phase, which suggests the compact object is located behind the Be disk, or surrounded by it.

S-shaped variations in Balmer lines reported by \citet{Aragona2010,Moritani2015} occurred after the primary outburst (phase of 0.55--0.95).
In the above geometry, this feature suggests that the disk perturbation is excited when the compact object passes nearby the Be star at the outburst.
Our monitoring revealed that emission lines have been stable in the last four years, which implies a stable Be disk.
In contrast, on the time-scale of the orbital period, the emission lines are variable; they change significantly after the primary outburst, and get stable around apastron.
The detailed profile variation in one orbital cycle will be discussed in another paper.

Recently, \citet{Yudin2017} reported multi-wavelength polarization at two phases in one orbital cycle, which showed variation in both polarization degree and polarization angle.
They discussed that the variability is possibly related with additional source or disk perturbation which occurs at periastron.
According to the orbital phase by \citet{Casares2012}, one observation was after periastron (0.039, 2015 March) and the other before it (0.859, 2015 December).
If the new ephemeris is adopted, these observational dates correspond to after apastron (0.37 for P=313 days, 2015 March), and before the successive apastron (0.99, 2015 December).
This implies that, even in the new orbit, the polalization is thought to vary when the compact object passes nearby the dense disk at the periastron.

%% --- Figure : New orbit --- %%
%% --- Figure : Mass function (end) --- %%

The new mass function (Figure \ref{fig:massf}) suggests that the mass of the compact object is $\leq 2.5 \; \mathrm{M_{sun}}$.
In the case of the orbital period of 313 days, the mass function and hence the mass of the compact object is even smaller.
The mass is thought to be below the Tolman-Oppenheimer-Volkoff limit ($3 \; \mathrm{M_{sun}}$) unless the inclination is low ($i < 3 ^{\circ}$), which suggests that $\mathrm{HESS~J0632+057}$ contains a neutron star.

\citet{Massi2017} have studied the relationship between $\Gamma$ and the X-ray luminosity $L_\mathrm{X}$, comparing with black hole X-ray binaries.
$\Gamma$ and $L_\mathrm{X}$ show a positive correlation in $\mathrm{PSR~B1259-63}$, while $\Gamma$ and $L_\mathrm{X}$ are anti-correlated in the X-ray black hole binaries \citep{Yang2015a,Yang2015b}.
In $\mathrm{HESS~J0632+057}$, the photon index $\Gamma$ is higher in the bright event than faint event.
This suggests that $\mathrm{HESS~J0632+057}$ is a system similar to $\mathrm{PSR~B1259-63}$.

\section{Conclusions} \label{sec:concl}
Optical monitoring of $\mathrm{HESS~J0632+057}$ for four years has successfully covered almost the whole orbital phase.
The radial velocity of H$\alpha$ emission line has been derived as bisector velocity of emission wing, which minimizes the effect of variation due to Be star activity \citep{Shafter1986}.
The resultant velocity variation suggests a different orbit from that proposed by \citet{Casares2012}.

Seven-year Swift/XRT data have been analyzed to search for the orbital modulation.
Using the X-ray light curve, the orbital period has been updated to be 313 days.
This is slightly shorter than previous determinations but fully consistent within errorbars.
The column density is higher around the outbursts.

The new orbit is smaller and still eccentric ($e \sim 0.6$), although lower eccentricity is suggested.
According to the new orbit, where the Be disk is easier to reach the orbit of the compact object at the phases other than periastron, X-ray activity and optical variation can be interpreted simply in the framework of the interaction between a misaligned Be disk and a compact object. 

The mass function indicates that the compact object is a $\leq 2.5 \; \mathrm{M_{sun}}$ neutron star.
The X-ray spectra tends to be softer when the source is brighter.
These facts suggest the X-ray emissions originate from the colliding system.
Further monitoring with good statistics are needed for confirming this scenario.

\begin{ack}
We deeply thank the anonymous referee for constructive comments and suggestions.
We are also grateful to Atsuo T. Okazaki for fruitful discussions.
A.~C.~C. acknowledges the support from CNPq (grant 307594/2015-7) and FAPESP (grant 2015/17967-7).
\end{ack}

\appendix

%%%
% See the manual for the detail.
%%%

\end{document}